\documentclass[10pt,letterpaper]{article}
\usepackage[
  a4paper, mag=1000, includefoot,
  left=3cm, right=1cm, top=2cm, bottom=2cm, headsep=1cm, footskip=1cm
]{geometry}
\usepackage{cite}
\usepackage{amssymb}
\usepackage{amsmath}

\usepackage{graphicx}
\graphicspath{{figures/}}
\usepackage{color}

\begin{document}

\begin{flushleft}
{\Large
\textbf\newline{Interaction of ultracold non-ideal ion-electron plasma with a uniform magnetic field}
}
\newline
\\
I. L. Isaev$^{1}$,
A. P. Gavriliuk$^{1,2}$,
\\
\bigskip
$^{1}$ Institute of Computational Modeling, Russian Academy of Sciences, Krasnoyarsk, Russia, 660036
\\
$^{2}$ Siberian Federal University, Krasnoyarsk, Russia, 660028
\end{flushleft}

\section*{Abstract}
The method of molecular dynamics is used to study behavior of  a ultracold non-ideal ion-electron Be$^+$ plasma in a uniform magnetic field. Our simulations yield an estimate for the rate of electron-ion collisions which is non-monotonically dependent on the magnetic field magnitude. Also they explicitly show that there are two types of diffusion: classical one, corresponding to Brownian motion of particles, and Bohm diffusion when the trajectory of particles (guiding centers) includes substantial lengths of drift motion.
\\
\\
PACS numbers: 52.27.Gr, 52.25.Xz


\section{Introduction}
\label{sec:intro}

Interest to systems of this kind is both fundamental and applied. Fundamental aspects, for instance, include feasibility to study kinetics of phase transitions in Coulomb systems as well as their crystallization (formation of a Wigner crystal). Such phase transitions have been observed in such Coulomb systems as dusty plasma~\cite{khrapak_04}. First attempts to obtain unltracold strongly non-ideal electron-ion plasma date back to 1999~\cite{PhysRevLett.83.4776,PhysRevLett.85.318}. However, the efforts to observe crystallization of this plasma failed because of its expansion and intrinsic processes in the plasma (disorder-induced heating~\cite{Zwicknagel1999,Murillo2001}, electron-ion energy exchange) resulting in heating of ions. To overcome this hindrance, laser cooling of ions is required, as shown earlier in~\cite{gav_97, Gavrilyuk1998}. Laser cooling techniques make it possible to produce an electron-ion plasma (free from expansion) with a strongly non-ideal ion subsystem having an ordered structure~\cite{Pohl2005,Killian2007,Zhang2008,PHR2009}. The major challenge is to make this plasma last long enough ($\geqslant 100$~µs) to allow ions to cool and an ordered structure to form. It would be natural to expect a magnetic field to substantially slow down plasma expansion. Indeed, in~\cite{Zhang2008} plasma expansion was reported to slow down under a weak uniform magnetic field. It should be noted that the available theoretical apparatus for describing plasma expansion (diffusion) and electron-ion energy exchange in a magnetic field~\cite{Silin1998} can as well be applied to weakly non-ideal systems of charged particles.

So, seeking a solution to the posed problem is motivated not only by studies of phase transitions in Coulomb systems but also by studies of behavior of non-ideal Coulomb systems in a magnetic field.

To the best of our knowledge, there have been no reports on simulation of behavior of an ultracold electron-ion plasma with a strongly non-ideal component in a magnetic field. Therefore realization of the model can be substantially useful for studying the properties of such plasma and finding possible ways to sustain it.

\section{Problem definition}
\label{sec:defination}

We consider a cylindrical column of ultracold electron-ion plasma aligned along a uniform magnetic field with induction $B$.

\begin{figure}[!ht]
\centerline{\includegraphics[width=50mm]{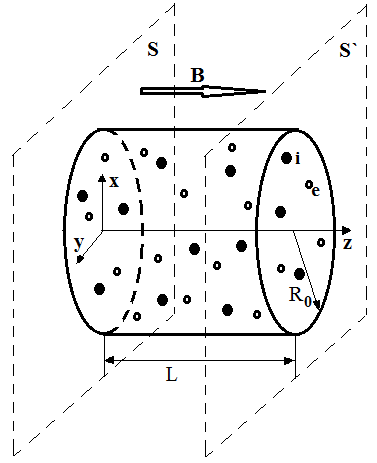}}
\caption{\label{fig:1} Sketched initial cell and initial distribution of ultracold plasma: $R_{0}$ is the initial cell radius, $L$ is the cell length, S and S' are elastic walls.}
\end{figure}

The molecular dynamics method was employed to study plasma dynamics using Be$^+$ plasma as an example.
The method based on the equation of motion for each particle:
\begin{equation} \label{eq0_}
\begin{gathered}
{{m}_{e}}\frac{d\mathbf{V}_{i}^{e}}{dt}=\sum\limits_{j\ne i}^{N}{\frac{{{e}^{2}}}{4\pi {{\varepsilon }_{0}}}\cdot \frac{(\mathbf{r}_{i}^{e}-\mathbf{r}_{j}^{e})}{{{\left| \mathbf{r}_{i}^{e}-\mathbf{r}_{j}^{e} \right|}^{3}}}}-\sum\limits_{k=1}^{N}{\frac{{{e}^{2}}}{4\pi {{\varepsilon }_{0}}}\cdot \frac{(\mathbf{r}_{i}^{e}-\mathbf{r}_{k}^{i})}{{{\left| \mathbf{r}_{i}^{e}-\mathbf{r}_{k}^{i} \right|}^{3}}}}-e[\mathbf{V}_{i}^{e}\times \mathbf{B}],\\
M\frac{d\mathbf{V}_{i}^{i}}{dt}=\sum\limits_{j\ne i}^{N}{\frac{{{e}^{2}}}{4\pi {{\varepsilon }_{0}}}\cdot \frac{(\mathbf{r}_{i}^{i}-\mathbf{r}_{j}^{i})}{{{\left| \mathbf{r}_{i}^{i}-\mathbf{r}_{j}^{i} \right|}^{3}}}}-\sum\limits_{k=1}^{N}{\frac{{{e}^{2}}}{4\pi {{\varepsilon }_{0}}}\cdot \frac{(\mathbf{r}_{i}^{i}-\mathbf{r}_{k}^{e})}{{{\left| \mathbf{r}_{i}^{i}-\mathbf{r}_{k}^{e} \right|}^{3}}}+}\,e[\mathbf{V}_{i}^{i}\times \mathbf{B}],
\end{gathered}
\end{equation}
where $m_e$ and $M$ are the masses of electron and ion, respectively.
$V_i^{e}$, $r_i^e$ are velocity and radius vector of the i-th electron.
$V_i^{i}$, $r_i^i$ are velocity and radius vector of the i-th ion.
$e$~---~elementary charge, $\varepsilon_0$~---~electric constant, $B$~---~vector of magnetic induction, $N$~---~number of particles (electrons or ions) in the cell.
The first two terms in the right-hand sides of equations \eqref{eq0_} describe the Coulomb interaction between charged particles, and the third describes the Lorentz force from the side of the external magnetic field.

Particles motion equations \eqref{eq0_} were solved by means of the Runge--Kutta--Fehlberg method of 4th order.
Computation was carried out for Be$^+$ plasma with the number of particles (ions, electrons) $N_{e} =N_{i} =N=200\div 1000$ for concentrations $n=10^{12} \div 10^{13} \, {\rm m}^{{\rm -3}} $ and magnetic field induction $B=0.01\div 0.1\, {\rm T}$.
The starting temperature of ions in all cases was $T_{i0} =0.01\, {\rm K}$ while the starting temperature of electrons varied within the range $T_{e0} =50\div 200\, {\rm K}$ and did not change further on.
That is, for the given concentrations, electrons can be considered weakly non-ideal (the electron non-ideality parameter is ${{\Gamma}_{e}}={{e}^{2}}/(4\pi {{\varepsilon }_{0}}a{{k}_{B}}{{T}_{e}})=0.0014\div 0.011$).

\section{Quasi-two-dimensional case}
\label{sec:quasi2d}

\begin{figure}[!ht]
\centerline{\includegraphics[width=130mm]{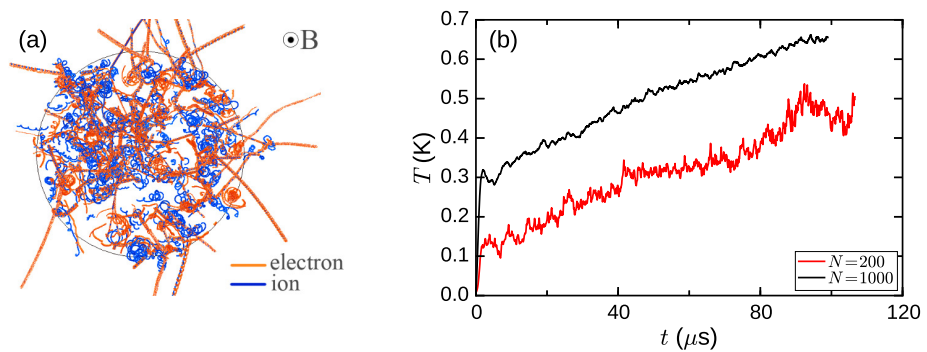}}
\caption{\label{fig:2} Trajectories of ions and electrons in a uniform magnetic field ($B=0.1\, {\rm T}$, $N=200$, $n=5\cdot 10^{12} \, {\rm m}^{{\rm -3}}$, $t=3\cdot 10^{-5} \, {\rm s}$, ${\rm R}_{0} =8.4\cdot 10^{-4} \, {\rm m}$) and (b) ion temperature for the same parameters (for $N=200,\, \, N=1000$).}
\end{figure}

Various formulations of the problem were considered. Initially we assumed walls $S$ and $S`$ to be elastic for both electrons and ions.
The number of particles varied within the range $N=200\div 1000$. Fig.~\ref{fig:2} shows typical behavior of particles when $L<a$, $a$ being the mean interparticle distance ($\frac{4}{3} \pi a^{3} n=1$), for $a=3.63\cdot 10^{-5} \, {\rm m}$  and $L\approx 1.8\cdot 10^{-5} \, {\rm m}$.
As seen from the figure, there are neutral electron-ion pairs formed in plasma, which travel freely across the magnetic field.
Movement of each individual particle of such a pair can be represented as motion in crossed magnetic and electric fields of an adjacent particle.
Possible existence of similar neutral complexes was shown earlier in~\cite{Kuzmin2004, Kuzmin2005} but, unlike those reports, localization of electrons along the magnetic field in our case is due to the elastic walls.

From Fig.~\ref{fig:2} one can find the typical plasma expansion velocity (for the initial plasma cloud radius $R_{0} =8.4\cdot 10^{-4} \, {\rm m}$)  $V_{dr} =15\div 20\, {\rm m/s}$. Alternatively, this drift velocity can be estimated as follows:

\begin{equation} \label{eq1_}
V_{dr} =\frac{E_{ei} }{B} ,\quad E_{ei} \approx \frac{e}{4\pi \varepsilon _{0} a_{ei}^{2} },
\end{equation}
where $a_{ei}$ is the mean distance between an electron and an ion (or between any two charges) given by

\begin{equation} \label{eq2_}
\frac{4}{3} \pi a_{ei}^{3} \cdot 2n=1.
\end{equation}

From \eqref{eq1_}, for the case shown in Fig.~\ref{fig:2}, we get $V_{dr} =16\, {\rm m/s}$, which is in fair agreement with the estimate obtained from the figure itself.

Despite a somewhat artificial situation under consideration, formation of neutral complexes can be observed under three-particle recombination\cite{Glinsky1991}.
In that case, an ion will trap an electron (in a magnetic field) and they will move on as a pair, which is prone to collapse when the magnetic field is removed.
Our results and findings from~\cite{Glinsky1991,Kuzmin2004,Kuzmin2005} suggest that using the immobile-ion approximation when simulating three-particle recombination in a strong magnetic field~\cite{Menshikov1995} is not correct.
Even though initially immobile, an ion can be dragged by an electron to drift together across the magnetic field.
At the initial stage the temperature of ions drastically increases.
This is believed to happen due to formation of neutral pairs as well as to disorder-induced heating of ions.
A further heating of ions is mainly associated with elastic collisions with electrons.

A three-dimensional case requires larger $L$. Plasma expansion behavior changes in that case: the number of neutral pairs reduces (until their complete disappearance when $L>a$) and expansion becomes diffusive.
Another distinctive feature is an increased initial surge of ion temperature. As can be seen from Fig.~\ref{fig:2}, it almost doubles for $L=3a \approx 10^{-4} \, {\rm m}$ and virtually no further increase is observed with growing $L$.
We attribute this anomalous heating of ions to multiple Coulomb interaction of electron with an adjacent ion which takes place in the adopted formulation of the problem (the presence of elastic walls for electrons).
Indeed, the average electron velocity along the magnetic field for the given electron temperatures is $V_{e\parallel} \sim 10^{4} \, {\rm m/s}$, hence the time it takes to travel the distance $L\approx 3a$ is $\tau_{\parallel} \sim 10^{-8} \, {\rm s}$, which is substantially shorter than the ion displacement time $\tau _{i} =\omega _{i}^{-1} \approx 10^{-6} {\rm s}$, where $\omega _{i} $ is the ion plasma frequency.
That is, hundreds of electron-ion collisions happen within that time. Using standard periodic conditions for electron does not help the situation: again there will be multiple interaction of an ion with actually the same electron.
Therefore we have modified the conditions for electrons as follows.
Let the escape point of electron through the wall $S`$ (Fig.~\ref{fig:1}) be given by coordinates $r`$ and $j`$, where $r`$ is the distance from the cell axis and $j`$ is the angle determining location of the escape point on a circumference of radius $r`$.
At the instant of an electron escape through $S`$ another electron arrives through the opposite wall at the point $r = r`$ at the same velocity as the escaped electron but the corresponding angle $j$ is chosen randomly.
So, taking $r = r`$ enables us to describe expansion of an electron cloud and the random choice of $j$ prevents multiple interaction of an ion with an adjacent electron.
It should be noted that it is possible to do without this approach, provided $L$ exceeds the mean free path length of an electron along the magnetic field.
However, for the concentrations under discussion this is of order $0.01\, {\rm m}$ or even longer. Hence the number of particles should be $N>10^{5}$, which will require considerable computational time and effort.

\section{Three-dimensional case}
\label{sec:3d}

The approach described above has been employed to calculate various characteristics of plasma depending on the applied field $B$, concentration of particles and ion temperature.

\subsection{Expansion}

\begin{figure}[!ht]
\centerline{\includegraphics[width=130mm]{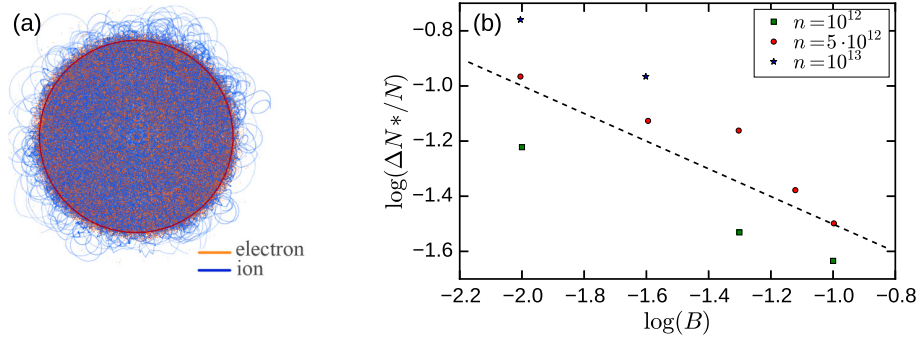}}
\caption{\label{fig:3} (a) Plasma cloud expansion across the magnetic field ($B=0.025\, {\rm T}$) beyond its initial domain (red circumference of radius $R_{0} =8.4\cdot 10^{-4} \; {\rm m}$) and (b) the relative fraction of electrons having escaped from the initial domain by the time $t=100\; {\rm \mu s}$ versus $B$ for $T_{e} =50\; {\rm K}$ and various concentrations.}
\end{figure}

An example of distribution of trajectories of ions and electrons across the magnetic field is given in Fig.~\ref{fig:3}(a) below for the concentration $n=5\cdot 10^{12} \, {\rm m}^{{\rm -3}}$, electron temperature $T_{e} =50\, {\rm K}$ and number of particles $N=1000$.
The mean Larmor radius of electrons $\rho_{eL}$ relates to the mean Larmor radius of ions $\rho _{eL}$ as

\begin{equation} \label{eq3_}
\frac{\rho _{eL}}{\rho_{iL}} =\frac{m_{e} V_{e\bot}}{MV_{i\bot}} =\left(\frac{m_{e} T_{e} }{MT_{i}} \right)^{1/2},
\end{equation}
where $T_{e} $ and $T_{i} $ are temperatures of electron and ion. Due to rapid heating of ions to $T_{i} >0.1\, {\rm K}$ (to be discussed later) the Larmor radius of electrons for $T_{e} <1500\, {\rm K}$ [as follows from \eqref{eq3_}] will be less than that of ions, so ions are the first to escape [see Fig.~\ref{fig:3}(a)].

In this case the limiting factor determining plasma expansion is escape of electrons. Our computational efforts were aimed at finding the relative fraction of electrons $\Delta N/N$ that appear outside the initial volume of plasma by the time $t$. In order to assess the fraction of electrons escaped due to expansion only $\Delta N^{*} /N$(ignoring those escaped due to Larmor motion), the following modification is to be made:

\begin{equation} \label{eq4_}
\frac{\Delta N^{*} }{N} =\frac{\Delta N}{N} -\frac{\rho _{eL} }{R_{0} }.
\end{equation}

Fig.~\ref{fig:3}(b) shows $\log (\Delta N^{*} /N)$ (obtained at $t=100\; {\rm \mu s}$) depending on $\log (B)$ for the electron temperature 50 K and various concentrations of particles. For comparison, also given there is the function (dashed line) $\log(1 / \sqrt{B})$, which, as can be seen, provides a good description of relative plasma expansion behavior depending on the magnetic field magnitude.

\subsection{Heating of ions by electrons}

\begin{figure}[!ht]
\centerline{\includegraphics[width=140mm]{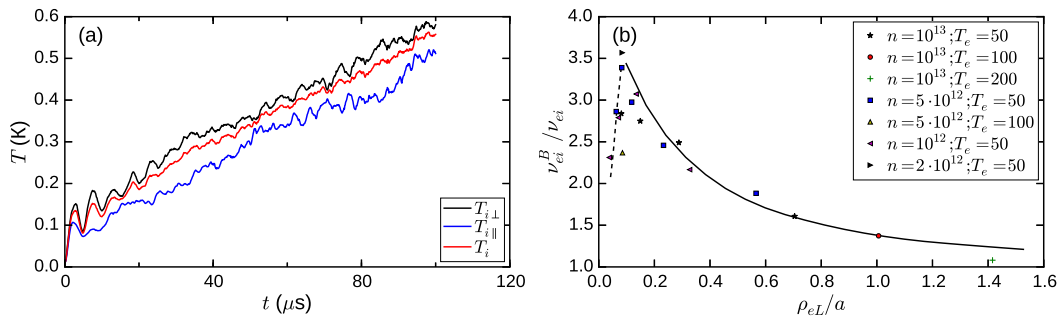}}
\caption{\label{fig:4} Ion temperature versus time (a) and collision rate ratio versus dimensionless electron Larmor radius (b).}
\end{figure}

Heating of ions by electrons during elastic collisions is a crucial process affecting behavior of ultracold non-isothermal plasma.
Considering $T_{e} \gg T_{i}$ and that electrons are weakly non-ideal, the heating will be defined as~\cite{Spitzer1962,PhysRevE.64.066411}

\begin{equation} \label{eq5_}
\frac{dT_{i}}{dt} \sim \frac{2m_{e}}{M} \nu _{ei} T_{e},
\end{equation}
where $\nu_{ei}$ is the electron-ion collisions rate dependent just on the electron velocity. Under a magnetic field, this rate ($\nu _{ei}^{B}$) will undergo change and can be deduced from dynamics of the ion temperature.
Fig.~\ref{fig:4}(a) illustrates time-dependent behavior of ion temperature for $n=5\cdot 10^{12} \; {\rm m}^{{\rm -3}} ,\quad T_{e} =50\; {\rm K},\quad B=0.1\; {\rm T}$.
As can be seen, the ion temperature is higher across the magnetic field than along the field ($T_{i\bot } >T_{i\parallel} $).
This is due to a faster heating in a cross-field direction and a relatively low rate of relaxation among the degrees of freedom (similar to the case considered for electrons in~\cite{Neil1985}).
The temperature difference drops with decreasing field or with growing concentration. In the following, we will use the average temperature $T_{i} = T_{i\bot } + 0.5T_{i\parallel}$ to find $\nu_{ei}^{B}$.
To avoid the influence of boundary effects, temperature estimates (as well as the diffusion coefficient) are obtained for particles in the central domain of the plasma cylinder of radius $r\le 0.75R_{0}$.

A temperature surge is observed at the initial stage though considerably weaker than in Fig.~\ref{fig:2}. We believe it to be associated with partial realization of disorder-induced heating with the account of Larmor motion.

The obtained dependence can be interpreted qualitatively as follows.
The increased $\rho_{eL} /a$ results from weakening of the magnetic field and its effect on electron-ion collisions, so $\nu _{ei}^{B} /\nu _{ei} \to 1$. In stronger fields, when $\rho _{eL} /a\to 0$, the Larmor radius is considerably shorter than the distance between particles and so the electric field of a rotating electron resembles that of a point charge moving along the  magnetic field.

\subsection{Diffusion of ions}

The issue of ion diffusion in ultracold electron-ion plasma is of undoubted interest.
There is a series of theoretical studies~\cite{Bernu1981,Marchetti1984,Ranganathan2003,Ott2011} where diffusion coefficients of ions under magnetic field are found using a one-component plasma (OCP) model.
But the OCP approximation is restricted as it ignores the discrete mode of interaction between ions and electrons and energy exchange between them as a result of collisions.

The cross-field diffusion coefficients are calculated by the Einstein relation for the mean-squared displacements~\cite{Hansen2006}:

\begingroup\makeatletter\def\f@size{8}\check@mathfonts
\begin{equation} \label{eq6_}
\begin{gathered}
D_{\bot} = \mathop{\lim }\limits_{\Delta t\to \infty } \frac{<\left|x(t+\Delta t)-x(t)\right|^{2} >+<\left|y(t+\Delta t)-y(t)\right|^{2} >}{4\Delta t},
\end{gathered}
\end{equation}
\endgroup

\begin{figure}[!ht]
\centerline{\includegraphics[width=70mm]{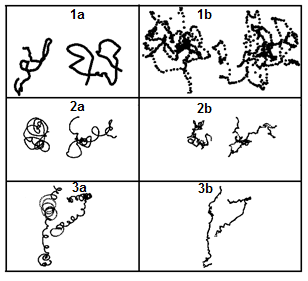}}
\caption{\label{fig:5} Trajectories of ions (a) and their guiding centers (b) for: 1 -- $B=0.01\; {\rm T}\; {\rm (}\beta {\rm =0.109)}$, 2 -- $B=0.05\; {\rm T}\; {\rm (}\beta {\rm =0.54)}$, 3 --$B=0.1\; {\rm T}\; {\rm (}\beta {\rm =1.09)}$.}
\end{figure}

Here $x$ and $y$ are the coordinates of ions at respective times, $<$\dots $>$ denotes averaging over all ensembles of atoms.
As for the diffusion coefficient along the magnetic field $B_{\parallel}$, it was not calculated because of the limited cell length along the z axis. The ion temperature grows due to electron-ion collisions; so the diffusion coefficient is bound to change as well. In order to reduce the effect of heating on the diffusion coefficient calculations, we have to restrict ourselves to a finite interval of $\Delta t=10^{-5} \, {\rm s}$,  where the condition $\Delta t>\omega _{i}^{-1} ,\; \omega _{iL}^{-1} $ is satisfied ($\omega _{i} ,\; \; \omega _{iL} $ is the plasma and cyclotron frequencys of ions) and the relative ion temperature change is not big. Note that the diffusion coefficient was calculated both from the coordinates of ions ($D_{\bot } $) and the coordinates of the guiding center $x_{c},\,y_{c} (D_{c\bot})$. However, the guiding center approximation is subject to certain conditions. We believe this approximation is applicable when the drift velocity of ion $V_{iidr}$, induced by the field of the nearest to it ion, is less than the mean thermal velocity $\left\langle V_{i\bot} \right\rangle$ in cross direction:

\begin{equation} \label{eq7_}
\begin{gathered}
V_{iidr} < \left\langle V_{i\bot }\right\rangle, \quad V_{iidr} =\frac{E_{ii} }{B} ,\quad E_{ii} \approx \frac{e}{4\pi \varepsilon _{0} a^{2} } \; \to \beta >0.44\Gamma _{i}^{1/2},
\end{gathered}
\end{equation}
where $\beta =\omega_{iL} /\omega_{i}$~\cite{Ott2011}, and $\Gamma_{i}={{e}^{2}}/(4\pi {{\varepsilon }_{0}}a{{k}_{B}}{{T}_{i}})$ is the nonideality parameter of the ion subsystem. Figure.~\ref{fig:5} shows trajectories (as projected on the xy plane) of two arbitrary ions and their guiding centers for various fields and concentration $n=5\cdot 10^{12} \, {\rm m}^{{\rm -3}}$ within the time interval 100 µs. In situations depicted in Fig.~\ref{fig:5}, the nonideality parameter $\Gamma _{i} $ changes approximately within the range 1$\div$4. In the case 1a, 1b condition \eqref{eq7_} is not applicable as it would lead to unreasonably large trajectory deviations. The case 2a and 2b is intermediate and deviations of trajectories of ions and their guiding centers are comparable. The same can be said of the case in Fig.~\ref{fig:5} (3a and 3b).

It can be seen from Fig.~\ref{fig:5} that the motion pattern of particles changes. In the cases 1ab and 2ab it is similar to Brownian motion of particles, which corresponds to classical diffusion~\cite{Lifshitz_82} $D_{\bot} \sim B^{-2} $.  In the latter case [Fig.~\ref{fig:5} (3a, 3b)], drift motion is observed, which suggests that we deal with Bohmian diffusion~\cite{Spitzer1960}. Based on the results obtained it can be stated that similar drift motion (Bohm diffusion) sets in when the mean Larmor radius of ions becomes less than the interparticle distance:

\begin{equation} \label{eq8}
\rho_{iL} \le a\, \text{or}\, \beta \ge 0.75\cdot \Gamma _{i}^{-1/2}.
\end{equation}

Trying to match this condition to the trajectories shown in Fig.~\ref{fig:5} reveals that it can be satisfied only in the case 3a, 3b.

\begin{figure}[!ht]
\centerline{\includegraphics[width=140mm]{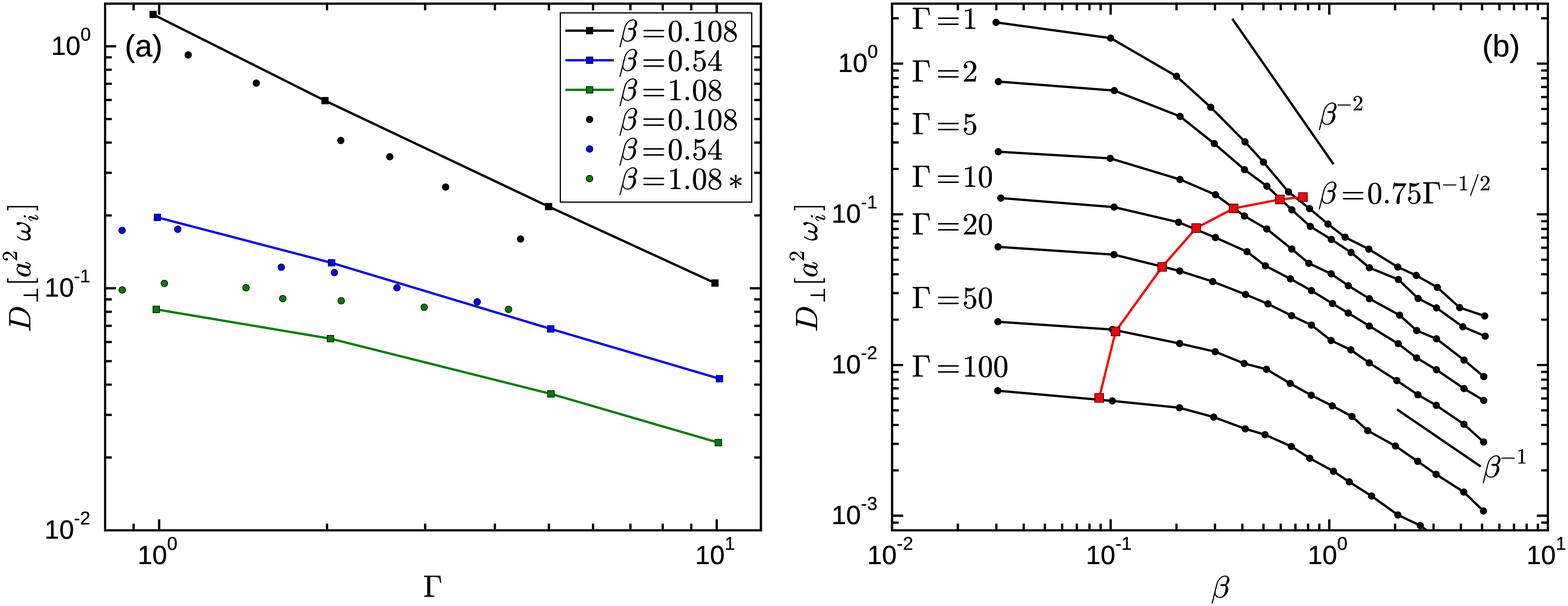}}
\caption{\label{fig:6} Cross diffusion coefficients depending on $\Gamma_{i} $(a) for various $\beta$, borrowed from~\cite{Ott2011} (squares) and obtained by us (circles) from particles or guiding centers (*). In the dependence of the diffusion coefficient on $\beta$ (b) for various $\Gamma_{i}$, borrowed again from~\cite{Ott2011}, also shown is the boundary of Bohm diffusion (solid red line) obtained from condition \eqref{eq8}.}
\end{figure}

Our estimates of diffusion coefficients were compared with the coefficients of cross diffusion from~\cite{Ott2011}. Some of the results of comparison (corresponding to the cases in Fig.~\ref{fig:5}) are illustrated in Fig.~\ref{fig:6}. As one can see, under classical diffusion conditions there is good agreement with OCP-based calculation results~\cite{Ott2011}. Furthermore, our findings show that changing the electron temperature from 50K to 200K has virtually no effect on the diffusion coefficient ($<$5\%). Thus, electrons present as discrete particles have very little effect on the ion diffusion coefficient. With regards to our results obtained under Bohm diffusion [Fig.~\ref{fig:5} (3a, 3b) and Fig.~\ref{fig:6} for $\beta =1.08$], the time interval, within which we calculate the diffusion coefficient, is too small in the presence of drift motion. In our opinion, the results from~\cite{Ott2011} are more adequate for the case.

In Fig.~\ref{fig:6}(b) we included a plot from~\cite{Ott2011} showing the boundary set by condition \eqref{eq8}. As follows from the figure, this boundary adequately describes the Bohm diffusion domain, where $D_{\bot} \sim B^{-1} $.

\section{Conclusions}
\label{sec:concl}

The obtained results lead us to conclude that in a quasi-two-dimensional case under fairly large magnetic fields ($B>0.01\; {\rm T}$) the process of formation of electron-ion pairs in a low-density ultracold plasma can give a noticeable contribution to the plasma heating and expansion.
It should be noted that this effect (formation of electron-ion pairs) is not observed in the 3D case.
The same process can also play a role in three-particle recombination giving rise to formation of weak electron-ion pairs~\cite{Kuzmin2004,Kuzmin2005}.

Adequate modeling of three-dimensional plasma using periodic conditions is feasible for large number of particles ($N>10^{5}$) when the mean free path of electron exceeds the cell length. To simplify the situation, we suggest the problem definition where electrons flow freely through the cell. We believe this makes an adequate modeling possible with a relatively small number of particles ($N>10^{3}$). Our calculations have yielded an estimate for the rate of elastic collisions leading to ion heating. This quantity, which is higher in a magnetic field than without it, is non-monotonically dependent on the magnetic field magnitude.

At moderate electron temperatures ($T_{e} \le 1000\, {\rm K}$) the rate of plasma expansion depends on the Larmor electron radius and at the initial stage $\sim B^{-1/2} $. The main hindrance to finding cross-field ion diffusion coefficients is associated with the change in the ion temperature (and consequently in the diffusion coefficients) due to electron-induced heating. Our simulations explicitly show that there are two types of diffusion: classical one, corresponding to Brownian motion of particles, and Bohmian diffusion when the trajectory of particles (guiding centers) includes substantial lengths of drift motion. In the former case the calculated diffusion coefficients are in good agreement with the results from~\cite{Ott2011} for the OCP model. This finding suggests that using the OCP model to calculate diffusion coefficients will be adequate for electron-ion plasma as well.

\clearpage
\bibliographystyle{ieeetr}

\begin{thebibliography}{10}

\bibitem{khrapak_04}
A.~G. Khrapak, S.~A. Khrapak, O.~F. Petrov, V.~Fortov, and V.~I. Molotkov {\em
  Phys. Usp.}, vol.~47, p.~447, 2004.

\bibitem{PhysRevLett.83.4776}
T.~C. Killian, S.~Kulin, S.~D. Bergeson, L.~A. Orozco, C.~Orzel, and S.~L.
  Rolston, ``Creation of an ultracold neutral plasma,'' {\em Phys. Rev. Lett.},
  vol.~83, pp.~4776--4779, Dec 1999.

\bibitem{PhysRevLett.85.318}
S.~Kulin, T.~C. Killian, S.~D. Bergeson, and S.~L. Rolston, ``Plasma
  oscillations and expansion of an ultracold neutral plasma,'' {\em Phys. Rev.
  Lett.}, vol.~85, pp.~318--321, Jul 2000.

\bibitem{Zwicknagel1999}
G.~Zwicknagel {\em Contrib.Plasma Phys.}, vol.~39, pp.~155--158, 1999.

\bibitem{Murillo2001}
M.~S. Murillo {\em Phys. Rev. Lett.}, vol.~87, p.~115003, 2001.

\bibitem{gav_97}
A.~P. Gavriliuk, I.~V. Krasnov, and N.~Y. Shaparev {\em Tech. Phys. Lett.},
  vol.~23, p.~61, 1997.

\bibitem{Gavrilyuk1998}
A.~P. Gavrilyuk, I.~V. Krasnov, and N.~Y. Shaparev, ``Laser cooling and wigner
  crystallization of resonant plasma in magnetooptical trap,'' {\em Laser
  Phys.}, vol.~8, no.~3, pp.~653--657, 1998.

\bibitem{Pohl2005}
T.~Pohl, T.~Pattard, and J.~M. Rost, ``Influence of electron–ion collisions
  on coulomb crystallization of ultracold neutral plasmas,'' {\em J. Phys. B:
  Atom. Mol. Opt. Phys}, vol.~38, pp.~343--350, Jan 2005.

\bibitem{Killian2007}
T.~C. Killian, T.~Pattard, T.~Pohl, and J.~M. Rost, ``Ultracold neutral
  plasma,'' {\em Physics Reports}, vol.~449, pp.~77--130, 2007.

\bibitem{Zhang2008}
X.~L. Zhang, R.~S. Fletcher, S.~L. Rolston, P.~N. Guzdar, and M.~Swisdak,
  ``Ultracold plasma expansion in a magnetic field,'' {\em Phys. Rev. Lett.},
  vol.~100, p.~235002, June 2008.

\bibitem{PHR2009}
A.~P. Gavriliuk, I.~L. Isaev, S.~V. Karpov, I.~V. Krasnov, and N.~Y. Shaparev,
  ``Brownian dynamics of laser cooling and crystallization of electron-ion
  plasma,'' {\em Physical Review E.}, vol.~80, pp.~054401--1--054401--6, 2009.

\bibitem{Silin1998}
V.~P. Silin, {\em Introduction to the Kinetic Theory of Gases}.
\newblock Moscow: Nauka, 1998.

\bibitem{Kuzmin2004}
S.~G. Kuzmin and T.~M. O'Neil {\em Phys. Plasmas}, vol.~11, p.~2382, 2004.

\bibitem{Kuzmin2005}
S.~G. Kuzmin and T.~M. O'Neil {\em Phys. Plasmas}, vol.~12, p.~012101, 2005.

\bibitem{Glinsky1991}
M.~E. Glinsky and T.~M. O'Neil {\em Physics of Fluids B: Plasma Physics},
  vol.~3, p.~1279, 1991.

\bibitem{Menshikov1995}
L.~I. Men'shikov and P.~. Fedichev {\em JETP}, vol.~81, no.~1, p.~78, 1995.

\bibitem{Spitzer1962}
J.~L. Spitzer, {\em Physics of Fully Ionized Gases}.
\newblock New-York-London: John Wiley and Sons Inc, 1962.

\bibitem{PhysRevE.64.066411}
G.~Hazak, Z.~Zinamon, Y.~Rosenfeld, and M.~W.~C. Dharma-wardana, ``Temperature
  relaxation in two-temperature states of dense electron-ion systems,'' {\em
  Phys. Rev. E}, vol.~64, p.~066411, Nov 2001.

\bibitem{Neil1985}
T.~M. O`Neil and P.~G. Hjorth {\em Phys. Fluids}, vol.~28, p.~3241, 1985.

\bibitem{Bernu1981}
B.~Bernu {\em J. Phys. Lett.}, vol.~42, p.~253, 1981.

\bibitem{Marchetti1984}
M.~C. Marchetti, T.~R. Kirkpatrick, and J.~R. Dorfman {\em Phys.Rev. A},
  vol.~29, p.~2960, 1984.

\bibitem{Ranganathan2003}
S.~Ranganathan, R.~Jonson, and C.~Woodward {\em Phys. Chem. Liq.}, vol.~41,
  p.~123, 2003.

\bibitem{Ott2011}
T.~Ott and M.~Bonitz {\em Phys. Rev. Lett.}, vol.~107, p.~135003, 2011.

\bibitem{Hansen2006}
J.~Hansen and I.~McDonald, {\em Theory of Simple Liquids}.
\newblock London: Academic, 2006.

\bibitem{Lifshitz_82}
E.~M. Lifshitz and L.~P. Pitaevski, {\em {Physical Kinetics}}.
\newblock New York: Pergamon, 1982.

\bibitem{Spitzer1960}
L.~Spitzer {\em Phys. Fluids}, vol.~3, p.~659, 1960.

\end{thebibliography}

\end{document}